\documentclass[aps,prd,showpacs,nofootinbib,tightenlines]{revtex4}
\usepackage[colorlinks,urlcolor=blue,linkcolor=blue,citecolor=blue]{hyperref} %
\usepackage{amsmath,amssymb,epsfig,graphicx,float,color,booktabs,ulem,verbatim,times} %
\usepackage[utf8]{inputenc}
\begin{document}
\title{\Large{The virtual contributions from $K^*$ in the $B \to K^*\pi h$ and $B \to K\rho h$ decays }\vspace{0.4cm}}

\author{Ai-Jun Ma$^{1}$}\email{theoma@163.com}
\author{Wen-Fei Wang$^{2}$}

\affiliation{\small
                     $^1$School of Mathematics and Physics, Nanjing Institute of Technology,
                            Nanjing, Jiangsu 211167, China \\
                    $^2$Institute of Theoretical Physics and State Key Laboratory of Quantum Optics and Quantum Optics Devices,
                            Shanxi University, Taiyuan, Shanxi 030006, China \\}

\date{\today}

\begin{abstract}
Inspired by the significant virtual contributions of the subprocesses $D^{*}/B^{*}\to D\pi$ and $ \rho\to\omega \pi/K\bar{K}$
in the three-body hadronic $B$ meson decays, we study the off-shell effects of the resonance $K^*$ decaying into the
$K^*\pi$ or $K\rho$ system in the decays $B \to K^*\pi h$ and $B \to K\rho h$ (with $h=\pi, K$) within the perturbative
QCD approach. The strong coupling constants $g_{K^*K^*\pi}$ and $g_{K^*K\rho}$ involved in this work are derived
from the coupling constant $g_{\rho\omega\pi}$ under the flavor SU$(3)$ symmetry. The $CP$ averaged branching fractions
for the quasi-two-body decays $B\to K^*[\to K^*\pi] h$ and $B\to K^*[\to K\rho] h$ are predicted to be on the order of $10^{-9}$
to $10^{-7}$ in this study. And the branching fractions for the decays $B\to K^*[\to K\rho] h$ are found to be around
 half of the corresponding results for the $B\to K^*[\to K^*\pi] h$ channels, mainly due to the phase space difference between
the $K\rho$ and $K^*\pi$ pairs originating from the intermediate state $K^*$. Experimental data for the $K^*$ decaying into the
final states $K^*\pi$ and $K\rho$ in the $B$ meson decays is expected to be obtained by the LHCb and Belle-II experiments
in the near future.
\end{abstract}


\maketitle


\section{Introduction}

Charmless three-body hadronic $B$ meson decays have attracted significant interest in recent years due to the valuable
$CP$ violation information and abundant resonance structures in their decay products.
Research on relevant decay processes imposes strong constraints on the Cabibbo-Kobayashi-Maskawa (CKM) matrix
elements~\cite{prl10-531,ptp49-652}, and could potentially provide evidence for physics beyond the Standard Model.
Utilizing the Dalitz-plot technique~\cite{PMS44-1068}, numerous experimental analyses for the three-body $B$ decays
have been performed by the {\it BABAR}~\cite{prd72-072003,prd74-032003,prd78-012004,prd78-052005,prd79-072006,
prd80-112001,prd85-112010,prd96-072001}, Belle~\cite{prd71-092003,prd75-012006,prl96-251803,prd79-072004}, and LHCb
Collaborations~\cite{prl111-101801,prl112-011801,prd90-112004,prl123-231802,prd101-012006,prl120-261801,prd108-012008}.
The rich phenomenology of $CP$ violation in these decays offers the opportunities to understand the complex interactions among
final states and probe the underlying $CP$ violation mechanism.
In particular, the observations of large localized $CP$ asymmetries have triggered extensive theoretical
investigations~\cite{prd74-114009,plb737-201,prd110-033001,plb726-337,plb727-136,prd89-074043,prd89-094013,prd92-094010,1512.09284,
plb806-135490,plb824-136824,prd88-114014,prd89-074025,prd94-094015,prd102-053006,plb728-579,plb561-258,prd91-014029,prd89-074031,
npb899-247,jhep10-117,epjc78-845,prd99-076010,epjc83-345,prd106-113002}.
Moreover, the outcomes of Dalitz-plot analyses typically include the contributions from various resonant states,
expressed in terms of the quasi-two-body fit fractions and $CP$ asymmetries.
In parallel, corresponding theoretical studies have been carried out within quasi-two-body framework-based methods
~\cite{prd79-094005,plb699-102,APPB42-2013,prd110-013002,prd76-094006,
plb763-29,prd95-056008,epjc79-37,jhep03-162,prd103-056021,epjc80-815,prd105-033003,epjc80-394,epjc80-517,prd110-036015,
prd111-053009,prd107-116023,epjc82-113,epjc84-753,prd108-016007}.

The resonance contributions in the three-body $B$ decays are predominantly localized around their respective pole
masses in the invariant mass distributions of the daughter particles. While the virtual contributions, also variously
known as the off-shell effects of resonances, are indispensable to the total amplitudes of specific decay channels,
even when the pole masses of the resonances lie far outside the kinematically allowed phase space.
For instance, the virtual contributions from the off-shell $D^*$ and $B^*$ in the $D\pi$ system were found
to be non-negligible in the amplitude analyses of $B \to D\pi h$ (with $h=\pi,K$)
decays~\cite{prd69-112002,prd76-012006,prd79-112004,prd91-092002,prd92-032002,prd94-072001}.
Specifically, the central values of the fit fractions for the virtual $D^{*0}$ and $B^{*0}$ states
were reported as $7.6\%$ and $3.6\%$ in the $B^- \to D^+\pi^-K^-$ decay~\cite{prd91-092002},
and $10.79\%$ and $2.69\%$ in the $B^- \to D^+\pi^-\pi^-$ decay~\cite{prd94-072001}, respectively.
The detailed discussion for the virtual contributions originated from $D^{*0}$ and
$D^{*\pm}$ in $B \to D\pi h$ decays on the theoretical side is found in Ref.~\cite{plb791-342}.
In Ref.~\cite{prl123-231802}, the LHCb Collaboration reported an unexpectedly large fit fraction around $30\%$ for
the subprocess $\rho(1450)\to K^+K^-$ in the $B^{\pm} \to \pi^{\pm} K^{+} K^{-}$ decay.
One of the best interpretations is that the contribution from the off-shell effect of $\rho(770)$ was included in the
measurement of the $\rho(1450)$ contribution for the kaon pair~\cite{prd101-111901}.
The results in Refs.~\cite{prd103-016002,prd103-056021,cpc46-053104} reveal that the quasi-two-body branching ratios
from the off-shell $\rho(770)$ for the kaon pair system in the relevant $B$ decays are comparable to those from the
excited states $\rho(1450)$ and $\rho(1700)$. In Ref.~\cite{prd92-012013},  the branching fraction of
$\bar B^0 \to D^{*+}\rho(770)^- [ \to \omega \pi^-]$ was measured to be $(1.48^{+0.37}_{-0.63}) \times 10^{-3}$,
which is larger than half of the total branching fraction $(2.46\pm0.18)\times 10^{-3}$ for the decay
$\bar B^0 \to D^{*+} \omega \pi^-$~\cite{PDG2024}, which means that a large coupling for $\rho(770) \to \omega\pi$
will significantly contribute to the distribution of $\omega \pi$ in the low-energy region for the related $B$ decays~\cite{jhep01-047}.

Motivated by the studies for the off-shell effects from $D^{*}/B^{*}\to D\pi$ and $\rho\to\omega \pi/K\bar{K}$
in the three-body $B$ decays, we will investigate in this work the virtual contributions arising from the subprocesses
$K^* \to K^*\pi$ and $K^* \to K\rho$ in the charmless three-body decays $B \to K^*\pi h$ and $B \to K\rho h$.\footnote{
For simplicity, we generally use the abbreviation $K^*=K^*(892)$ and $\rho=\rho(770)$ in the following sections.}
As is well known, the $K^*$ resonance predominantly decays to its final state $K\pi$, with a branching fraction close to
$100\%$~\cite{PDG2024}. The natural decay modes $K^* \to K^*\pi$ and $K^*\to K \rho$ are kinematically forbidden,
as the $K^*$ pole mass lies below the thresholds of its decay products.
However, the $K^*K^*\pi$ and $K^*K\rho$ vertices are playing the crucial role in the theoretical studies
of the $K^*\to K\pi\pi$ decay rate~\cite{prd10-2993,lnc18-83,plb66-165,epjc39-205}.
Given the large available phase space, the contributions from the off-shell $K^*$ could be substantial in relevant
three-body $B$ meson decays.
For the $B \to K^*\pi h$ and $B \to K\rho h$ decays, the experimental measurements are found for
$B^+ \to K^{*+}\pi^-\pi^+$, $B^+ \to K^{*+}\pi^\mp K^\pm$, $B^0 \to K^{*0}\pi^+\pi^-$ and $B^0 \to K^{*0}\pi^\pm K^\mp$ by the
ARGUS~\cite{plb262-148}, {\it BABAR}~\cite{prd74-051104,prd76-071104}, and Belle Collaborations~\cite{prd80-051103,prd81-071101}.
Meanwhile, Belle has reported the first observation of the three-body decay $B^0 \to K^+\rho^0\pi^-$~\cite{prd80-051103}.
The total branching fractions (or upper limits) for the concerned three-body decays were measured, but the corresponding
partial branching fractions for the decays involving $K^*\pi$ and $ K\rho$ pairs have yet to be determined in these experiments.
Within the factorization approach, the branching ratios for the decays $B^+ \to K^{*+}\pi^+\pi^-$ and $B^0 \to K^{*0}\pi^+\pi^-$
were calculated~\cite{epja50-122}, considering only resonance contributions for the $\pi^+\pi^-$ system.

In the previous studies~\cite{plb763-29,prd95-056008,epjc79-37,jhep03-162,prd103-056021,epjc80-815,prd105-033003,epjc80-394,
epjc80-517,prd110-036015,prd111-053009,epjc84-753}, the contributions from multiple intermediate resonance states in the charmless
three-body $B$ meson decays have been investigated by using the perturbative QCD (PQCD) approach~\cite{prd63-054008,plb504-6,prd63-074009}
based on the $k_T$ factorization theorem. The consistency between the theoretical predictions and experimental data
shows the applicability of the PQCD factorization approach in studying the quasi-two-body $B$ meson decays.
In this work, we shall focus on the virtual contributions from $K^*$ in the $B \to K^*\pi h$ and $B \to K\rho h$ decays
within the framework of the PQCD approach.
The excitations $K^*(1410)$ and $K^*(1680)$ have been observed to decay to $K^*\pi$ and $K\rho$
final states in the reaction $K^-p \to \bar{K}^0 \pi^+\pi^-n$ by the LASS Collaboration~\cite{plb149-258,npb292-693},
with the branching ratios $\mathcal{B}(K^*(1410) \to K^*\pi(K\rho))>40\%(<7\%)$ and
$\mathcal{B}(K^*(1680) \to K^*\pi(K\rho))=29.9^{+2.2}_{-5.0}\%(31.4^{+5.0}_{-2.1}\%)$~\cite{PDG2024}.
Evidences for $K^*(1410,1680)\to K^*\pi$ or $K\rho$ in the $K\pi\pi$ distribution have also been found in the amplitude
analyses of $B^+ \to J/\psi(\psi(2S)) K^+\pi^+\pi^-$~\cite{prd83-032005,jhep01-054} and $B^+ \to \chi_{c1(2)} K^+\pi^+\pi^-$
decays~\cite{prd93-052016}.
Since well-defined distribution amplitudes for higher-mass $K^*$ excitations and  precise experimental data are currently lacking,
we leave the contributions from $K^*(1410,1680)\to K^*\pi$ and $K\rho$ in the concerned decays to future studies.

The rest of this paper is organized as follows. In Sec.~\ref{sec-frame}, we give a brief review of the PQCD framework
for the concerned decay processes. In Sec.~\ref{sec-result}, we present the numerical results and give necessary discussions.
A summary is presented in Sec.~\ref{sec-sum}. The Lorentz-invariant decay amplitudes and the factorization formulas in the PQCD approach
for the relevant decays are provided in the Appendix~\ref{sec-appx-a}.
\section{Framework}
\label{sec-frame}
\begin{figure}[tbp]
\centerline{\epsfxsize=14cm \epsffile{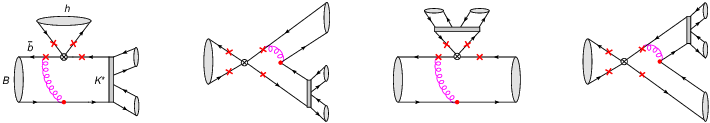}}
\caption{The leading order Feynman diagrams for the decays $B\to K^* h$ with the subprocesses $K^*\to K^*\pi$ and $K^*\to K\rho$.
         The symbol $\otimes$ represents the insertion of the four-fermion vertices in the effective theory and $\times$ denotes
          the possible attachments of hard gluons.}
\label{fig-feyndiag}
\end{figure}
In the PQCD formalism, the decay amplitude for the hadronic $B$ meson decay is expressed as the convolution of a hard kernel with
the distribution amplitudes of both the initial $B$ meson and the final-state hadrons.
For a quasi-two-body decay $B\to R[\to h_1h_2]h$, the $h_1h_2$ pair which proceeds by the intermediate state $R$ moves collinear
while the bachelor meson $h$ recoils energetically.
Compared with the framework for two-body $B$ decays, the two-meson distribution amplitudes are newly introduced to describe the
interaction between the meson pair $h_1h_2$.
Then, the factorization formula of the decay amplitude ${\mathcal A}$ for the $B\to K^*[\to K^*\pi(K\rho)]h$ decay is written as
~\cite{plb561-258,plb763-29}
\begin{eqnarray}
{\mathcal A}=\phi_B \otimes {\mathcal H} \otimes  \phi^{P\text{-wave}}_{K^*\pi(K\rho)} \otimes \phi_{h},
\label{def-DA-Q2B}
\end{eqnarray}
where the symbol $\otimes$ stands for the convolution in parton momenta and the hard kernel ${\mathcal H}$ is given by
the Feynman diagrams with one hard gluon exchange.
The complete set of leading-order Feynman diagrams for the concerned decays is presented in Fig.~\ref{fig-feyndiag}.
Among these, Figs.~\ref{fig-feyndiag}(a) and~\ref{fig-feyndiag}(c) are referred to as the emission diagrams,
while Figs.~\ref{fig-feyndiag}(b) and~\ref{fig-feyndiag}(d) represent the annihilation diagrams.
As essential inputs, the universal hadron distribution amplitudes $\phi_B$, $\phi^{P\text{-wave}}_{K^*\pi(K\rho)}$, and $\phi_h$
absorb the nonperturbative dynamics in the hadronization processes.
In this work, we employ the light-cone distribution amplitudes for the heavy $B$ meson and the light $\pi$ and $K$ mesons
as specified in Refs.~\cite{jhep03-162,prd103-056021,prd76-074018}, with detailed functional forms and parameters provided
therein and in the references they contain.

  For the decay process of a vector resonance ($V_R$) into a vector ($V$) and a pseudoscalar ($P$) meson,
  the related form factor $F_{VP}(s)$ can been defined from the matrix element~\cite{epja38-311,prd92-014014,prd108-074025}
  \begin{eqnarray}
      \langle V (p_V,\lambda) P(p_P)| j_\mu(0) | 0 \rangle
            = i  \epsilon_{\mu\nu\alpha\beta}\, \varepsilon^{\nu}(p_V, \lambda)\, p_P^\alpha \, p^\beta F_{VP}(s)\, ,
  \label{strongmatrix}
  \end{eqnarray}
  where $j_\mu$ is the isovector part of the electromagnetic current, $\varepsilon$ and $\lambda$ are the polarization vector
  and polarization parameter for the final state $V$, and $p=p_V+p_P$ and $s$ are the momentum and the squared invariant mass for the resonance $V_R$, respectively.
  In the vector meson dominance model, the form factor $F_{VP}(s)$, which incorporates the coupling for $V_R \to VP$, is parametrized as
  ~\cite{plb486-29,prd94-112001,PPNP120-103884,prd88-054013}
 \begin{eqnarray}
	F_{VP}(s) = \frac{g_{V_RVP}}{g_{V_R}}\frac{m_{V_R}^2 }{D_{V_R}(s)} \, .
  \label{emformfactor}
 \end{eqnarray}
   Here, the excited states of $V_R$ in the form factor have not been taken into account.
  Then, the decay amplitude for the quasi-two-body process $B\to K^*[\to K^*\pi(K\rho)] h$
  can be written as
\begin{equation}
     {\mathcal M}^{Q2B} = \langle(K^*\pi(K\rho))_{K^*} h |{\mathcal H}_{\rm eff}|B\rangle=g_{K^*K^*\pi(K\rho)}\epsilon_{\mu\nu\alpha\beta}
                          \varepsilon^\mu_{K^*_R}\varepsilon^{\nu}_{K^*(\rho)} p_{\pi(K)}^\alpha p^\beta
                            \frac{1}{D_{K^*}(s)}{\mathcal M}^{2B} \varepsilon^*_{K^*_R}\cdot p_B,
                               \label{def-Op-Q2B}
\end{equation}
  where ${\mathcal H}_{\rm eff}$ represents the effective Hamiltonian and ${\mathcal M}^{2B} \varepsilon^*_{K^*_R}\cdot p_B$ is the decay amplitude for the $B \to K^*h$ decay.
  To make a distinction, $\varepsilon_{K^*_R}$ denotes the polarization vector for the intermediate state $K^*$, and $\varepsilon_{K^*}$ is for the final state $K^*$.
  In this work, the components of the amplitude associated with the form factor are parametrized into the distribution amplitudes
  for $K^* \to K^*\pi(K\rho)$ in the final form of amplitude ${\mathcal A}$ for the quasi-two-body decay, while the parts of polarizations and momenta are
  included in the formula of the differential branching fraction as shown in Eq.~(\ref{eqn-diff-bra}).

For the $P$-wave distribution amplitudes of the $K^*\pi$ ($K\rho$) system, we here only need to consider the longitudinal
polarization component, which can be collected into~\cite{prd76-074018}
\begin{eqnarray}
  \phi^{P\text{-wave}}_{K^*\pi(K\rho),L}(x,s)&=&\frac{-1}{\sqrt{6}}
      \big[\sqrt{s}\,{\epsilon\hspace{-1.5truemm}/}\!_L\phi^0(x,s)
             + {\epsilon\hspace{-1.5truemm}/}\!_L {p\hspace{-1.7truemm}/} \phi^t(x,s)
           + \sqrt s \phi^s(x,s)  \big]\!,
\end{eqnarray}
with $p$ and $s$ denoting the momentum and the square of invariant mass for the $K^*\pi(K\rho)$ pair, respectively.
The symbol $\epsilon_L$ represents the longitudinal polarization vector and $x$ is the momentum fraction of the
positive quark in the intermediate vector resonance $K^*$.
The twist-$2$ and twist-$3$ distribution amplitudes are parametrized as
\begin{eqnarray}
   \phi^{0}(x,s)&=&\frac{3f_{K^*\pi(K\rho)}(s)}{\sqrt{6}}x(1-x)\big[1+a_{1K^*}^{||}3(2x-1)+a_{2K^*}^{||}\frac{3}{2}(5(2x-1)^2-1)\big],\label{def-DA-0}\\
    \phi^{t}(x,s)&=&\frac{3f_{K^*\pi(K\rho)}^T(s)}{2\sqrt{6}}(2x-1)^2,\label{def-DA-t}\\
   \phi^{s}(x,s)&=&\frac{3f_{K^*\pi(K\rho)}^T(s)}{2\sqrt{6}}(1-2x),\label{def-DA-s}
\end{eqnarray}
with the Gegenbauer moments $a_{1K^*}^{||}=0.03\pm0.02$ and $a_{2K^*}^{||}=0.11\pm0.09$~\cite{prd76-074018}.
The factor $f_{K^*\pi(K\rho)}(s)$ in Eq.~(\ref{def-DA-0}) is related to the transition form factor
$F_{K^*\pi(K\rho)}(s)$, which includes the coupling for $K^*\to K^*\pi(K\rho)$.
Following the treatment of the subprocess $\rho\to\omega \pi$ in Ref.~\cite{jhep01-047}, we have the parametrization
\begin{eqnarray}
	f_{K^*\pi(K\rho)}(s) = \frac{ g_{K^* K^*\pi(K\rho)}m_{K^*}f_{K^*}}{D_{K^*}(s)}\,,
  \label{exp-formfactor}
\end{eqnarray}
where $m_{K^*}$ and $f_{K^*}$ denote the mass and decay constant of the $K^*$ meson, respectively.
Here, the denominator $D_{K^*}$, modeled by a Breit-Wigner propagator, is defined as
\begin{equation}
    D_{K^*}(s) = m_{K^*}^2-s-i \sqrt s\,\Gamma_{K^*}(s).
  \label{BW-denom}
\end{equation}
Following the form for $\rho\to\omega\pi$ in Refs.~\cite{prd88-054013,prd108-092012}, the energy-dependent width of the $K^*$
including the coupled channels $K\pi$, $K^*\pi$, and $K\rho$ can be expressed as
\begin{eqnarray}
	\Gamma_{K^*}(s) &=& \Gamma_{K^*}\frac{m^2_{K^*}}{s}
		\left(\frac{q_1(s)} {q_1(m^2_{K^*})}\right)^{3}
		+\frac{g_{K^*K^*\pi}^2}{12\pi} q^3_2(s)+\frac{g_{K^*K\rho}^2}{12\pi} q^3_3(s),
\label{eq-Gm-R892}
\end{eqnarray}
where the momentum magnitudes of one of the decay products in the rest frame of the intermediate state are given by
\begin{eqnarray}
        q_1(s) =\frac{\sqrt{\left[s-(m_K+m_{\pi})^2\right]\left[s-(m_K-m_{\pi})^2\right]}}{2\sqrt s} , \label{def-q1}   \qquad
\end{eqnarray}
 \begin{eqnarray}
     q_2 (s)=\frac{\sqrt{\left[s-(m_{K^*}+m_{\pi})^2\right]\left[s-(m_{K^*}-m_{\pi})^2\right]}}{2\sqrt s} ,\label{def-q2}\qquad
\end{eqnarray}
and
 \begin{eqnarray}
     q_3 (s)=\frac{\sqrt{\left[s-(m_K+m_{\rho})^2\right]\left[s-(m_K-m_{\rho})^2\right]}}{2\sqrt s},\label{def-q3}\qquad
\end{eqnarray}
respectively.
In this work, the coupling constants $g_{K^*K^*\pi}$ and $g_{K^*K\rho}$ are derived through the flavor SU(3) symmetry relation,
explicitly expressed as $g^2_{K^*K^*\pi}=g^2_{K^*K\rho}=\frac{3}{4}g^2_{\rho\omega\pi}$~\cite{jhep07-276,prd96-054033}.
According to the discussions in Ref.~\cite{jhep01-047}, we adopt the input value $g_{\rho\omega\pi}=16.0\pm2.0$ GeV$^{-1}$.
The derived coupling $g^2_{K^*K^*\pi}$ is consistent with the result $\frac{g^2_{K^*K^*\pi}}{4\pi}=18$ GeV$^{-2}$
in Ref.~\cite{lnc18-83}.
We also employ isospin symmetry, taking $K^{*+}$ as an example, which leads to the relation
$|g_{K^{*+}K^{*0}\pi^+(K^0\rho^+)}|=\sqrt2|g_{K^{*+}K^{*+}\pi^0(K^+\rho^0)}|=\frac{\sqrt 6}{3}|g_{K^*K^*\pi(K\rho)}|$.
For the factor $f_{K^*\pi(K\rho)}^T(s)$ in the twist-$3$ distribution amplitudes $\phi^{t,s}(x,s)$, we use the assumption
$f_{K^*\pi(K\rho)}^T(s)/f_{K^*\pi(K\rho)}(s)\approx f_{K^*}^T/f_{K^*}$ as in~\cite{plb763-29} with the result
$ f_{K^*}^T/f_{K^*}=0.853$~\cite{prd76-074018}.

With the help of the polarization sum relations
\begin{eqnarray}
      \sum_{\lambda=0,\pm} \varepsilon^{*\mu}{(p,\lambda)}\varepsilon^{\nu}{(p,\lambda)}=-g^{\mu\nu}
            +\frac{p^\mu p^\nu}{p^2}, \label{polarization1}
\end{eqnarray}
\begin{eqnarray}
      \sum_{\lambda=0,\pm}\vert\epsilon_{\mu\nu\alpha\beta}p_h^\mu\varepsilon^{\nu}(p_{V}, \lambda)
          p_P^\alpha p^\beta\vert^2  %
            = s\,\vert\overrightarrow{p_V}\vert^2 \vert\overrightarrow{p_h}\vert^2 (1-\cos^2{\theta}), \label{polarization2}
\end{eqnarray}
the differential branching fraction of the $B\to K^*[\to K^*\pi(K\rho)]h$ decays can be expressed as~\cite{jhep01-047}
\begin{eqnarray}
 \frac{d{\mathcal B}}{d s}=\tau_B\frac{sq_{2(3)}^3q_h^3}{24\pi^3m^7_B}{|{\mathcal A}|^2},
    \label{eqn-diff-bra}
\end{eqnarray}
$\mathcal{\tau}_B$ being the mean lifetime of the $B$ meson. In the rest frame of the intermediate resonance, the magnitudes of the
momenta $q_2$ and $q_3$ are related to the $K^*\pi$ and $K\rho$ pairs as shown in Eqs.~(\ref{def-q2}) and (\ref{def-q3}),
and the expression of $q_h$ for the bachelor meson $h$ is described as
\begin{eqnarray}
   q_h (s)=\frac{\sqrt{\left[m^2_{B}-(\sqrt s+m_{h})^2\right]\left[m^2_{B}-(\sqrt s-m_{h})^2\right]}}{2\sqrt s},\quad
 \label{eq-qh}
\end{eqnarray}
where $m_B$ and $m_{h}$ are the masses for the $B$ meson and the bachelor $h$, respectively.
By combining contributions from the relevant Feynman diagrams shown in Fig.~\ref{fig-feyndiag}, we obtain the Lorentz-invariant
decay amplitudes ${\mathcal A}$ for the considered $B$ decays, which are provided in the Appendix~\ref{sec-appx-a}.
In addition, the $CP$ averaged branching ratio $\mathcal B$ is defined as
\begin{eqnarray}
{\mathcal B}=\frac{{\mathcal B}(\bar B\to \bar f)+{\mathcal B}(B\to f)}{2},
\end{eqnarray}
while the direct $CP$ asymmetry ${\mathcal A}_{CP}$ is given by
\begin{eqnarray}
{\mathcal A}_{CP}=\frac{{\mathcal B}(\bar B\to \bar f)-{\mathcal B}(B\to f)}{{\mathcal B}(\bar B\to \bar f)+{\mathcal B}(B\to f)},
\end{eqnarray}
where $f$ denotes the final state and $\bar B \to \bar f$ is the $CP$ conjugate mode of $B \to f$.

\section{Results and discussions}
\label{sec-result}
Before the numerical analysis, we first introduce the input parameters adopted in our calculations in Table.~\ref{tab:inputs},
which includes the masses and decay constants for the initial and final states, the decay widths of the $K^*$, the lifetimes of
the $B_{(s)}$ mesons, and the Wolfenstein parameters of the CKM matrix.
\begin{table}[H]
\caption{ The input parameters adopted in numerical calculations~\cite{prd76-074018,PDG2024}.}
\label{tab:inputs}
\centering
\begin{tabular*}{12cm}{@{\extracolsep{\fill}}llll}
  \hline\hline
  \text{Mass (\text{GeV})}
  &$m_{B^{\pm}}=5.279$  &$m_{B^0}=5.280$  &$m_{B^0_s}=5.367$   \\[1ex]
  &$m_{K^{*\pm}}=0.892$ &$m_{K^{*0}}=0.896$ &$m_\rho=0.775$ \\[1ex]
  &$m_{K^{\pm}}=0.494$ &$m_{K^0}=0.498$ & $m_{\pi^{\pm}}=0.140$\\[1ex]
  &$m_{\pi^0}=0.135$ && \\[1ex]\hline
  \text{Decay constant (\text{GeV})}
  &$f_{B}=0.190$& $f_{B_s}=0.230$  &$f_\pi=0.130$  \\[1ex]
  &$f_K=0.156$&$f_{K^*}=0.217$  &$f_{K^*}^T=0.185$   \\[1ex]\hline
  \text{Decay width (\text{GeV})}
  &$\Gamma_{K^{*\pm}}=0.0514$  &$\Gamma_{K^{*0}}=0.0473$  &   \\[1ex]\hline
  \text{Lifetime (ps)}
  &$\tau_{B^{\pm}}=1.638$  &$\tau_{B^0}=1.517$  &$\tau_{B^0_s}=1.520$  \\[1ex]\hline
  \text{Wolfenstein parameters}
  &$\lambda=0.22501\pm0.00068$  & ~~~~~~~~~~$A=0.826^{+0.016}_{-0.015}$ & \\[1ex]
  &$\bar{\rho}=0.1591\pm0.0094$ &  ~~~~~~~~~~$\bar{\eta}=0.3523^{+0.0073}_{-0.0071}$ & \\[1ex]
\hline\hline
\end{tabular*}
\end{table}

\begin{table}[thb]   
\begin{center}
\caption{PQCD predictions of the $CP$ averaged branching fractions and the direct $CP$
         asymmetries for the quasi-two-body $B\to K^*[\to K^*\pi] h$ decays.}
\label{Br1}
\begin{tabular}{l c c } \hline\hline
 ~~~~~~~~~~~~~Decay modes     &    $\mathcal B$    &  ${\mathcal A}_{CP}(\%)$            \\
\hline  %
  $B^+ \to K^{*0}[\to K^{*+}\pi^-] \pi^+$    &~~$(4.59^{+1.42+0.05+0.59}_{-1.00-0.06-0.57})\times 10^{-7}$~~~
      & ~~$-1.5^{+0.2+0.1+0.1}_{-0.3-0.3-0.2}$~~    \\
  $B^+ \to K^{*+}[\to K^{*0}\pi^+]\pi^0$     &~~$(3.41^{+1.07+0.06+0.32}_{-0.75-0.06-0.31})\times 10^{-7}$~~~
      & ~~$-21.2^{+2.7+2.2+1.1}_{-2.7-2.1-1.1}$~~     \\
  $B^+ \to K^{*+}[\to K^{*0}\pi^+]\bar K^0$    &~~$(1.50^{+0.34+0.13+0.37}_{-0.27-0.12-0.32})\times 10^{-8}$~~~
      & ~~$-34.4^{+2.2+2.7+6.4}_{-2.5-2.2-5.2}$~~    \\
  $B^+ \to \bar K^{*0}[\to K^{*-}\pi^+]K^+$    &~~$(3.04^{+0.94+0.15+0.39}_{-0.56-0.05-0.33})\times 10^{-8}$~~~
      & ~~$11.0^{+2.1+5.7+3.5}_{-2.0-5.8-2.4}$~~    \\
\hline %
  $B^0 \to K^{*+}[\to K^{*0}\pi^+]\pi^-$     &~~~$(3.69^{+1.05+0.12+0.46}_{-0.73-0.09-0.44})\times 10^{-7}$~~~
      & ~~$-42.0^{+4.6+4.6+2.2}_{-4.0-3.8-2.1}$~~    \\
  $B^0 \to K^{*0}[\to K^{*+}\pi^-]\pi^0$    &~~~$(1.02^{+0.24+0.05+0.19}_{-0.16-0.03-0.18})\times 10^{-7}$~~~
      & ~~$-9.4^{+0.3+1.2+1.1}_{-0.1-0.6-1.1}$~~    \\
  $B^0 \to K^{*+}[\to K^{*0}\pi^+]K^-$        &~~~$(6.89^{+0.51+0.34+2.35}_{-0.51-0.32-1.75})\times 10^{-9}$~~~
      & ~~$-32.0^{+3.3+2.8+4.4}_{-5.1-4.2-7.9}$~~    \\
  $B^0 \to K^{*-}[\to \bar K^{*0}\pi^-]K^+$         &~~~$(3.49^{+0.39+1.66+0.57}_{-0.22-1.15-0.39})\times 10^{-9}$~~~
      & ~~$19.8^{+0.5+6.9+9.6}_{-2.3-17.0-12.0}$~~    \\
  $B^0 \to K^{*0}[\to K^{*+}\pi^-]\bar K^0$    &~~~$(1.34^{+0.24+0.20+0.32}_{-0.18-0.18-0.27})\times 10^{-8}$~~~
      & ~~ 0 ~~    \\
  $B^0 \to \bar K^{*0}[\to K^{*-}\pi^+]K^0$     &~~~$(2.83^{+0.98+0.06+0.39}_{-0.68-0.05-0.37})\times 10^{-8}$~~~
      & ~~ 0 ~~    \\
\hline %
  $B_s^0 \to K^{*-}[\to \bar K^{*0}\pi^-]\pi^+$       &~~~$(7.09^{+2.56+0.60+0.06}_{-1.79-0.59-0.06})\times 10^{-7}$~~~
      & ~~$-13.9^{+1.9+1.5+1.2}_{-2.1-1.4-1.2}$~~    \\
  $B_s^0 \to \bar K^{*0}[\to K^{*-}\pi^+]\pi^0$     &~~~$(1.22^{+0.51+0.13+0.19}_{-0.33-0.12-0.18})\times 10^{-8}$~~~
      & ~~$-70.4^{+7.9+6.5+6.7}_{-7.5-6.0-6.1}$~~    \\
  $B_s^0 \to K^{*+}[\to K^{*0}\pi^+]K^-$          &~~~$(4.67^{+1.48+0.15+0.59}_{-0.98-0.10-0.56})\times 10^{-7}$~~~
      & ~~$-36.1^{+4.7+5.1+0.8}_{-3.7-4.4-0.7}$~~    \\
  $B_s^0 \to K^{*-}[\to \bar K^{*0}\pi^-]K^+$             &~~~$(5.53^{+1.19+0.72+0.80}_{-0.84-0.64-0.72})\times 10^{-7}$~~~
      & ~~$49.8^{+0.1+1.7+5.2}_{-0.8-2.7-4.9}$~~    \\
  $B_s^0 \to K^{*0}[\to K^{*+}\pi^-]\bar K^0$  &~~~$(5.73^{+1.83+0.11+0.73}_{-1.25-0.10-0.71})\times 10^{-7}$~~~
      & ~~ 0 ~~    \\
  $B_s^0 \to \bar K^{*0}[\to K^{*-}\pi^+]K^0$   &~~~$(3.59^{+0.49+0.70+0.69}_{-0.30-0.62-0.59})\times 10^{-7}$~~~
      & ~~ 0 ~~    \\
\hline\hline
\end{tabular}
\end{center}
\end{table}
\begin{table}[thb]   
\begin{center}
\caption{PQCD predictions of the $CP$ averaged branching fractions and the direct $CP$
         asymmetries for the quasi-two-body $B\to K^*[\to K\rho] h$ decays.}
\label{Br2}
\begin{tabular}{l c c } \hline\hline
 ~~~~~~~~~~~~~Decay modes     &    $\mathcal B$    &  ${\mathcal A}_{CP}(\%)$            \\
\hline  %
  $B^+ \to K^{*0}[\to K^+\rho^-] \pi^+$    &~~$(2.46^{+0.76+0.22+0.32}_{-0.53-0.22-0.30})\times 10^{-7}$~~~
      & ~~$-1.1^{+0.2+0.3+0.1}_{-0.2-0.4-0.1}$~~    \\
  $B^+ \to K^{*+}[\to K^0\rho^+]\pi^0$     &~~$(1.91^{+0.60+0.18+0.17}_{-0.43-0.18-0.17})\times 10^{-7}$~~~
      & ~~$-20.3^{+2.6+2.2+1.0}_{-2.5-2.0-1.0}$~~     \\
  $B^+ \to K^{*+}[\to K^0\rho^+]\bar K^0$    &~~$(9.37^{+2.40+1.28+2.48}_{-1.84-1.10-2.10})\times 10^{-9}$~~~
      & ~~$-29.8^{+2.8+1.5+4.6}_{-2.8-0.9-3.2}$~~    \\
  $B^+ \to \bar K^{*0}[\to K^-\rho^+]K^+$    &~~$(1.57^{+0.47+0.15+0.20}_{-0.30-0.13-0.16})\times 10^{-8}$~~~
      & ~~$5.2^{+3.0+7.5+4.5}_{-1.7-6.9-4.9}$~~    \\
\hline %
  $B^0 \to K^{*+}[\to K^0\rho^+]\pi^-$     &~~~$(2.02^{+0.57+0.19+0.25}_{-0.40-0.19-0.24})\times 10^{-7}$~~~
      & ~~$-41.5^{+4.4+4.6+2.0}_{-4.0-3.9-2.0}$~~    \\
  $B^0 \to K^{*0}[\to K^+\rho^-]\pi^0$    &~~~$(5.20^{+1.17+0.47+0.99}_{-0.79-0.45-0.92})\times 10^{-8}$~~~
      & ~~$-9.7^{+0.3+1.3+1.0}_{-0.1-0.9-1.1}$~~    \\
  $B^0 \to K^{*+}[\to K^0\rho^+]K^-$        &~~~$(3.48^{+0.24+0.28+1.01}_{-0.23-0.31-0.80})\times 10^{-9}$~~~
      & ~~$-46.0^{+3.6+2.3+4.3}_{-4.6-3.1-4.1}$~~    \\
  $B^0 \to K^{*-}[\to \bar K^0\rho^-]K^+$         &~~~$(2.01^{+0.19+0.92+0.30}_{-0.16-0.72-0.22})\times 10^{-9}$~~~
      & ~~$23.8^{+2.7+6.7+10.7}_{-4.0-9.9-12.8}$~~    \\
  $B^0 \to K^{*0}[\to K^+\rho^-]\bar K^0$    &~~~$(7.82^{+1.62+1.41+2.00}_{-1.22-1.21-1.65})\times 10^{-9}$~~~
      & ~~ 0 ~~    \\
  $B^0 \to \bar K^{*0}[\to K^-\rho^+]K^0$     &~~~$(1.59^{+0.54+0.16+0.22}_{-0.38-0.15-0.21})\times 10^{-8}$~~~
      & ~~ 0 ~~    \\
\hline %
  $B_s^0 \to K^{*-}[\to \bar K^0\rho^-]\pi^+$       &~~~$(3.68^{+1.33+0.41+0.03}_{-0.93-0.41-0.03})\times 10^{-7}$~~~
      & ~~$-12.8^{+1.8+1.5+1.1}_{-2.0-1.4-1.1}$~~    \\
  $B_s^0 \to \bar K^{*0}[\to K^-\rho^+]\pi^0$     &~~~$(7.74^{+3.41+1.12+1.09}_{-2.23-1.05-1.02})\times 10^{-9}$~~~
      & ~~$-59.8^{+7.6+6.7+6.2}_{-7.9-6.0-5.8}$~~    \\
  $B_s^0 \to K^{*+}[\to K^0\rho^+]K^-$          &~~~$(2.67^{+0.84+0.27+0.34}_{-0.57-0.26-0.32})\times 10^{-7}$~~~
      & ~~$-34.4^{+4.4+5.0+0.7}_{-3.8-4.3-0.5}$~~    \\
  $B_s^0 \to K^{*-}[\to \bar K^0\rho^-]K^+$             &~~~$(3.12^{+0.77+0.47+0.47}_{-0.53-0.43-0.42})\times 10^{-7}$~~~
      & ~~$47.1^{+0.5+2.0+5.2}_{-1.5-3.3-5.1}$~~    \\
  $B_s^0 \to K^{*0}[\to K^+\rho^-]\bar K^0$  &~~~$(3.21^{+1.03+0.32+0.41}_{-0.69-0.30-0.39})\times 10^{-7}$~~~
      & ~~ 0 ~~    \\
  $B_s^0 \to \bar K^{*0}[\to K^-\rho^+]K^0$   &~~~$(1.94^{+0.33+0.38+0.39}_{-0.21-0.35-0.34})\times 10^{-7}$~~~
      & ~~ 0 ~~    \\
\hline\hline
\end{tabular}
\end{center}
\end{table}
Based on the formula of the differential branching ratio in Eq.~(\ref{eqn-diff-bra}) and the decay amplitudes given in the Appendix~\ref{sec-appx-a},
we calculate the $CP$ averaged branching fractions and the direct $CP$ asymmetries for the quasi-two-body decays
$B\to K^*[\to K^*\pi] h$ and $B\to K^*[\to K\rho] h$, with the results listed in Tables~\ref{Br1} and~\ref{Br2}, respectively.
The first error for each PQCD prediction comes from the uncertainty of the shape parameter in the $B$ meson distribution amplitudes,
specifically $\omega_B=0.40\pm0.04~{\rm GeV}$ for $B^{+,0}$ or $\omega_{B_s}=0.50\pm0.05~{\rm GeV}$ for $B_s^0$.
The second error is induced by the Gegenbauer moments $a_{1K^*}^{||}=0.03\pm0.02$ and $a_{2K^*}^{||}=0.11\pm0.09$
for the $K^*\pi(K\rho)$ distribution amplitudes, combined with the coupling constant $g_{\rho\omega\pi}=16.0\pm2.0$ GeV$^{-1}$,
which is utilized to determine the values of $g_{K^*K^*\pi}$ and $g_{K^*K\rho}$.
The last one stems from the variations of the chiral masses $m^{\pi}_0=(1.4\pm0.1)~{\rm GeV}$ and $m^K_0=(1.6\pm0.1)~{\rm GeV}$,
as well as the Gegenbauer moment $a^h_2 = 0.25\pm0.15$ for the distribution amplitudes of bachelor pion and kaon.
The errors associated with uncertainties in other parameters are negligible and have been omitted here.
One interesting thing is that the error for the PQCD-predicted branching fraction $\mathcal{B}(B\to K^*[\to K\rho] h)$
caused by the uncertainty of $g_{\rho\omega\pi}$ adopted in this work is larger than that for the corresponding branching
fraction $\mathcal{B}(B\to K^*[\to K^*\pi] h)$.
This is because the coupling constant $g_{K^*K^*\pi(K\rho)}$ derived from $g_{\rho\omega\pi}$ appears in both the numerator and denominator of
the factor $f_{K^*\pi(K\rho)}(s)$ in Eq.~(\ref{exp-formfactor}). While this configuration partially cancels the variation effect of
$g_{K^*K^*\pi(K\rho)}$ on the branching fraction, the cancellation magnitude depends critically on the terms independent
of $g_{K^*K^*\pi(K\rho)}$ in the denominator.
Here, different ranges of the invariant mass squared $s$ in these two types of decay produce varying cancellation magnitudes.
Due to the isospin symmetry, the branching fractions of the quasi-two-body decays involving a neutral pion (or rho meson) as a decay product
of the intermediate $K^*$ state are half of those listed in Tables~\ref{Br1} and~\ref{Br2}, while their $CP$ asymmetries remain identical.

As shown in Tables~\ref{Br1} and~\ref{Br2}, the branching fractions contributed by the off-shell $K^*$ states for the considered
three-body decays range from $10^{-9}$ to $10^{-7}$. Among them, the branching fractions predicted for the decays $B\to K^*[\to K\rho] h$
in Table~\ref{Br2} are approximately $55\%$ of those for $B\to K^*[\to K^*\pi] h$ in Tables~\ref{Br1}.
This suppression arises from phase space differences; namely, the mass combination $m_K+m_\rho$ is greater than $m_{K^*}+m_\pi$.
In addition, one can observe similar $CP$ asymmetries for the same decay with different subprocesses in Tables~\ref{Br1} and~\ref{Br2}.
In the Standard Model, direct $CP$ violation in a $B$ meson decay arises from the interference of tree and penguin amplitudes.
The predicted large $CP$ asymmetries are attributed to sizable interference between these amplitudes, whereas small ones result from
scenarios where either the tree or penguin amplitude dominates. Since only the penguin diagrams contribute to the
$B_{(s)}^0 \to K^{*0}[\to K^{*+}\pi^-(K^+\rho^-)]\bar K^0$ and $B_{(s)}^0 \to \bar K^{*0}[\to K^{*-}\pi^+(K^-\rho^+)]K^0$ decays,
direct $CP$ violation does not occur in these modes.

\begin{table}[thb]   
\begin{center}
\caption{Available experimental data on the branching fractions and $CP$ asymmetries for the $B\to K^*h$ decays~\cite{PDG2024}.}
\label{Br3}
\begin{tabular}{c c c } \hline\hline
  ~~~~~Decay modes~~~~~     &    $\mathcal B_{exp}$    &  ${\mathcal A}_{CP}(\%)$            \\
\hline  %
  $B^+ \to K^{*0}\pi^+ $    &~~$(1.01\pm0.08)\times 10^{-5}$~~~
      & ~~$-2.1\pm3.2$~~    \\
  $B^+ \to K^{*+}\pi^0 $     &~~$(6.8\pm0.9)\times 10^{-6}$~~~
      & ~~$-39\pm21$~~     \\
  $B^+ \to \bar K^{*0} K^+ $    &~~$(5.9\pm0.8)\times 10^{-7}$~~~
      & ~~$4\pm5$~~    \\
\hline %
  $B^0 \to K^{*+}\pi^- $     &~~~$(7.5\pm0.4)\times 10^{-6}$~~~
      & ~~$-27\pm4$~~    \\
  $B^0 \to K^{*0}\pi^0 $    &~~~$(3.3\pm0.6)\times 10^{-6}$~~~
      & ~~$-15\pm13$~~    \\
  $B^0 \to K^{*\pm}K^\mp $        &~~~$<4\times 10^{-7}$~~~
      & ~~ ... ~~     \\
  $B^0 \to K^{*0}\bar K^0 +\bar K^{*0}K^0$     &~~~$<9.6\times 10^{-7}$~~~
      & ~~ ... ~~     \\
\hline %
  $B_s^0 \to K^{*-}\pi^+ $       &~~~$(2.9\pm1.1)\times 10^{-6}$~~~
      & ~~ ... ~~     \\
  $B_s^0 \to K^{*\pm}K^\mp $          &~~~$(1.9\pm0.5)\times 10^{-5}$~~~
      & ~~ ... ~~     \\
  $B_s^0 \to K^{*0}\bar K^0 +\bar K^{*0} K^0$  &~~~$(2.0\pm0.6)\times 10^{-5}$~~~
      & ~~ ... ~~    \\
\hline\hline
\end{tabular}
\end{center}
\end{table}

In Table~\ref{Br3}, we present the available experimental data of the branching fractions and $CP$ asymmetries for the two-body $B\to K^*h$
decays~\cite{PDG2024}. Under the isospin symmetry of strong decays, the subprocesses $K^* \to K\pi$ satisfy the relationships
$\frac{\mathcal B (K^* \to K \pi^\pm)}{\mathcal B (K^* \to K \pi)}= \frac{2}{3}$ and $\frac{\mathcal B (K^* \to K \pi^0)}{\mathcal B (K^* \to K \pi)}= \frac{1}{3}$.
By combining the data in Table~\ref{Br3} and assuming ${\mathcal B (K^*\to K\pi})\approx 100\%$, one can obtain the branching fractions
for the $B\to K^*[\to K\pi^\pm] h$ decays via the quasi-two-body approximation
\begin{eqnarray}
 {\mathcal B (B\to K^*[\to K\pi^\pm] h)}={\mathcal B (B\to K^* h)}\times {\mathcal B (K^*\to K\pi^\pm)}.
 \label{eq-KSKpi}
\end{eqnarray}
Defining the ratio
\begin{eqnarray}
 R_{K^*\pi}=\frac{{\mathcal B (B\to K^*[\to K^*\pi^\pm] h)}_{\rm PQCD}}{{\mathcal B (B\to K^*[\to K\pi^\pm] h)}_{\rm exp}},
 \label{eq-ratioKpi}
\end{eqnarray}
we find that $R_{K^*\pi}$ is concentrated between approximately $5\%$ and $8\%$ for all measured decay channels,
which is quite small as expected, except for the case related to $B_s^0 \to K^{*-}\pi^+$.
But it should be noted that a significant discrepancy persists between the theoretical predictions and experimental measurement of
${\mathcal B (B_s^0 \to K^{*-}\pi^+})$, with the predicted value~\cite{epjc79-37,prd76-074018,npb675-333,prd80-114026,prd91-014011,
prd91-074026,epjc85-544} being about three times larger than measured~\cite{PDG2024,NJP16-123001}.
A similar ratio $R_{K\rho}$ is also $55\%$ of $R_{K^*\pi}$.
Collectively, the $K^*\pi$ and $K\rho$ contributions are negligible and physically insignificant, considering that the $K^*$ predominantly
decays via the $K\pi$ channel $(\approx100\%)$. As the phase space expands in $B$ meson decays, however, the ratios of branching fractions
for associated quasi-two-body channels become significantly enhanced.

Up to now, several three-body decays of $B \to K^*\pi h$ have been observed by the ARGUS, {\it BABAR}, and Belle Collaborations
\cite{plb262-148,prd74-051104,prd76-071104,prd80-051103,prd81-071101}. In this work, the central values of the PQCD-predicted branching fractions
for the decays $B^+ \to K^{*0}[\to K^{*+}\pi^-] \pi^+$, $B^0 \to K^{*+}[\to K^{*0}\pi^+]\pi^-$, and $B^0 \to K^{*+}[\to K^{*0}\pi^+]K^-$ correspond to
$0.61 \%$, $0.67\%$, and $0.15\%$ of the measured three-body branching fractions ${\mathcal B}(B^+ \to K^{*+}\pi^- \pi^+)=(7.5\pm1.0) \times 10^{-5}$,
${\mathcal B}(B^0 \to  K^{*0}\pi^+\pi^-)=(5.5\pm0.5) \times 10^{-5}$, and ${\mathcal B}(B^0 \to K^{*0}\pi^+K^-)=(4.5\pm1.3) \times 10^{-6}$,
respectively, as presented in the {\it Review of Particle Physics}~\cite{PDG2024}, where these values represent averages derived from results
in Refs.~\cite{plb262-148,prd74-051104,prd76-071104,prd80-051103,prd81-071101}.
Besides, the $B^0 \to K^+\rho^0K^-$ decay has also been observed for the first time by the Belle Collaboration~\cite{prd80-051103} and
the corresponding partial branching fraction for $m_{K\pi} \in [0.75,1.20]~\rm GeV$ was also presented. Under isospin symmetry, one also has
${\mathcal B}(B^0 \to K^{*+}[\to K^+\rho^0]\pi^-)\approx \frac{{\mathcal B}(B^0 \to K^{*+}[\to K^0\rho^+]\pi^-)}{2}=1.01 \times 10^{-7}$,
which is $3.6\%$ of the measured ${\mathcal B}(B^0 \to K^+\rho^0\pi^-)=(2.8\pm0.7)\times 10^{-6}$ by Belle.
Compared to the $B \to D\omega \pi$ channel, where off-shell $\rho$ contribution dominates~\cite{prd92-012013,PDG2024,jhep01-047}, the PQCD
predictions indicate that the $K^* \to K^*\pi$ and $K\rho$ contributions in the concerned three-body $B$ decays are substantially smaller.
Similar conclusions were obtained in a very recent study for the charmed decays $B\to DV^*[\to VP]$~\cite{2506.14675}, but
they considered only the $K^*\to K\rho$ contribution and ignored the more important $K^*\to K^*\pi$ term in the specific decay processes.

\begin{figure}[thb]
\centerline{\epsfxsize=9 cm \epsffile{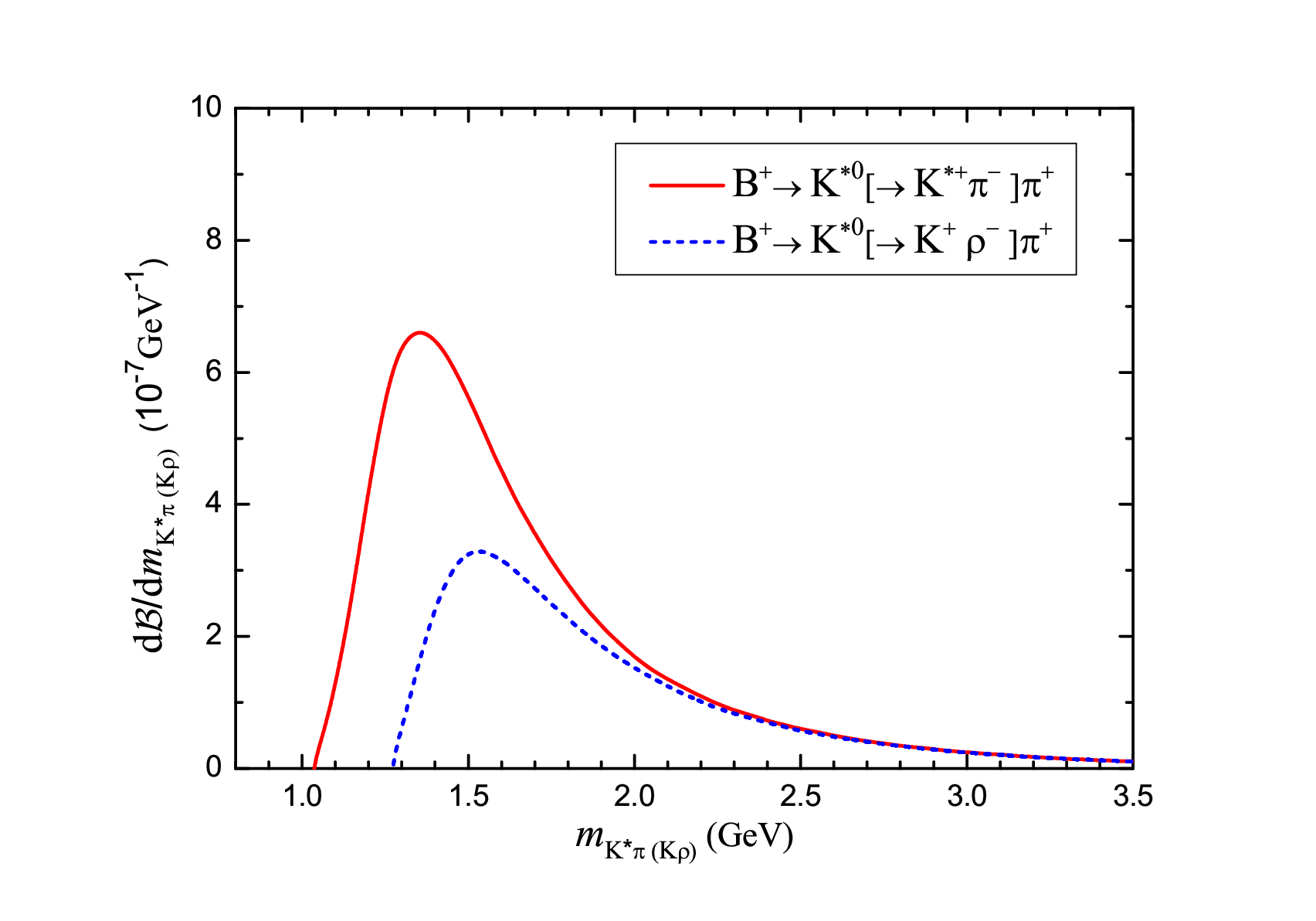}}
\caption{Differential branching fractions for the $B^+ \to K^{*0}[\to K^{*+}\pi^-] \pi^+$ and $B^+ \to K^{*0}[\to K^+\rho^-] \pi^+$
decays with the invariant mass $m_{K^*\pi}$ or $m_{K\rho}$ ranging from threshold to $3.5~\rm GeV$.}
\label{fig-diff}
\end{figure}

Unlike the two-body $B$-meson decays with fixed kinematics, the quasi-two-body decay amplitudes exhibit strong dependence on the invariant mass of intermediate
resonance daughter particles. In Fig.~\ref{fig-diff}, we plot the differential branching fractions for the $B^+ \to K^{*0}[\to K^{*+}\pi^-] \pi^+$ and
$B^+ \to K^{*0}[\to K^+\rho^-] \pi^+$ decays with the invariant mass $m_{K^*\pi(K\rho)}$ ranging from threshold to $3.5~\rm GeV$.
The area between each curve and the horizontal axis reflects the difference in branching ratios of the two decay channels as discussed in the previous paragraph.
For a full resonant contribution in the $B$ meson decay, the bulk of the corresponding branching fraction concentrates where the invariant mass of
the decay products equals the pole mass of the resonance.
However, strong phase-space suppression near threshold shifts the peak positions in the differential branching fractions for the $B^+ \to K^{*0}[\to K^{*+}\pi^-] \pi^+$
and $B^+ \to K^{*0}[\to K^+\rho^-] \pi^+$ decays to approximately $1.4~\rm GeV$ and $1.6~\rm GeV$; this feature could be misinterpreted as the signal of new resonance contributions.
Meanwhile, the off-shell components may also bias the extraction of contributions from excited states such as the $K^{*}(1410)$ and $K^{*}(1680)$.
It should also be noted that in the kinematic region of $K^*\pi$ or $K\rho$ where the virtual $K^*$ contribution exists, there may be contributions
from other full resonance states and final-state interactions. The predicted $CP$ asymmetries in this work can be clearly observed only if
the virtual contribution of the $K^*$ is dominant. However, due to the current lack of Dalitz plot analysis and partial wave analysis
for those three-body $B$ meson decays,  it is difficult to determine the significance of such virtual contributions in the discussed phase space,
necessitating further experimental verification.

\begin{table}[thb]   
\begin{center}
\caption{Available experimental data on the branching fractions for the $B\to K^*(1410,1680)h$ decays~\cite{PDG2024}.
 Upper limits for the branching fractions are set at 90\% confidence level.}
\label{Br4}
\begin{tabular}{l c  } \hline\hline
 ~~~~~~~~~~~~Decay modes     &    ~~~~~~~~~~~~${\mathcal B}_{exp}$             \\
\hline  %
  ~~~~$B^+ \to K^{*}(1410)^0\pi^+$    &~~~~~~~~~~~~$<4.5\times 10^{-5}$~~~~   \\
  ~~~~$B^+ \to K^{*}(1680)^0\pi^+$    &~~~~~~~~~~~~$<1.2\times 10^{-5}$~~~~   \\
\hline %
  ~~~~$B^0 \to K^{*}(1680)^0\pi^0$     &~~~~~~~~~~~~$<7.5\times 10^{-6}$~~~~   \\
\hline %
  ~~~~$B^0 \to K^{*}(1410)^+\pi^-$    &~~~~~~~~~~~~$<8.6\times 10^{-5}$~~~~  \\
  ~~~~$B^0 \to K^{*}(1680)^+\pi^-$    &~~~~~~~~~~~~$(1.41\pm0.10) \times 10^{-5}$~~~~   \\
\hline\hline
\end{tabular}
\end{center}
\end{table}

In Table~\ref{Br4}, we list the available data on the branching fractions for the $B\to K^*(1410,1680)h$ decays~\cite{PDG2024}, derived from experimental
analyses of the related quasi-two-body modes with the subprocesses $K^*(1410,1680) \to K\pi$~\cite{prd72-072003,prd78-052005,prd71-092003,prd75-012006,prl120-261801}.
Except for ${\mathcal B}(B^0 \to K^*(1680)^+\pi^-)$ which has a definite value, the others are only given an upper limit at $90\%$ confidence level.
Using the PDG values ${\mathcal B} (K^*(1680)\to K^*\pi)=29.9\%$ and ${\mathcal B} (K^*(1680)\to K\rho)=31.4\%$, together with $2/3$ for the $K^{*0}\pi^+$
and $K^0\rho^+$ fractions, one obtains
\begin{eqnarray}
 {\mathcal B}(B^0 \to K^{*}(1680)^+[\to K^{*0}\pi^+]\pi^-)&=&(2.81\pm0.20) \times 10^{-6},\\
 {\mathcal B}(B^0 \to K^{*}(1680)^+[\to K^0\rho^+]\pi^-)&=&(2.95\pm0.21) \times 10^{-6}.
    \label{bra1680}
\end{eqnarray}
Then, it is easy to obtain the ratios  $\frac{{\mathcal B}(B^0 \to K^{*+}[\to K^{*0}\pi^+]\pi^-)_{PQCD}}{{\mathcal B}(B^0 \to K^{*}(1680)^+[\to K^{*0}\pi^+]\pi^-)}= 13.1\%$
and $\frac{{\mathcal B}(B^0 \to K^{*+}[\to K^0\rho^+]\pi^-)_{PQCD}}{{\mathcal B}(B^0 \to K^{*}(1680)^+[\to K^0\rho^+]\pi^-)}=6.8\%$,
suggesting that the off-shell $K^*$ contribution is not very significant in the partial wave analysis of the $K^*$ family. It is important to notice that the reported
${\mathcal B}(B^0 \to K^*(1680)^+\pi^-)=(1.41\pm0.10) \times 10^{-5}$ by LHCb~\cite{prl120-261801} is consistent with {\it BABAR}'s result
${\mathcal B}(B^0 \to K^*(1680)^+\pi^-)<2.5 \times 10^{-5}$~\cite{prd78-052005} but not with ${\mathcal B}(B^0 \to K^*(1680)^+\pi^-)<1.0 \times 10^{-5}$
presented by Belle~\cite{prd75-012006}. Due to the lack of precise experimental data for other decays, this conclusion still needs to be further confirmed.
Namely, future high-statistics experiments by Belle II and LHCb are expected to search for the $K^{*}$ meson and its excited particles through intermediate states
$K^*\pi$ and $K\rho$ in the $K\pi\pi$ spectrum from relevant charmless $B$ meson decays.

\section{SUMMARY}
\label{sec-sum}
In this work, we studied the virtual contributions originated from $K^*\to K^*\pi$ and $K^*\to K\rho$ in the charmless three-body $B$ meson decays by employing the PQCD approach.
We calculated the $CP$ averaged branching fractions and the direct $CP$ asymmetries for the quasi-two-body $B\to K^*[\to K^*\pi] h$ and
$B\to K^*[\to K\rho] h$ decays. The branching fractions for the concerned decays range from $10^{-9}$ to $10^{-7}$, among which the predicted
${\mathcal B} (B\to K^*[\to K\rho] h)$ are approximately half of ${\mathcal B} (B\to K^*[\to K^*\pi] h)$.
The natural decay modes of $K^* \to K^* \pi$ and $K\rho$ are blocked since the resonance pole mass is below the thresholds of the $K^* \pi$ and $K\rho$ pairs,
and the $K^*$ meson is widely accepted as decaying almost one hundred percent into $K\pi$. As the phase space expands in $B$ meson decays, however, the ratios
of the predicted ${\mathcal B} (B\to K^*[\to K^*\pi^\pm] h)$ to the experimentally measured ${\mathcal B} (B\to K^*[\to K\pi^\pm] h)$ mainly range from $5\%$ to $8\%$ in this work.
Several decays of $B \to K^*\pi h$ and $B \to K\rho h$ have been measured by different collaborations but without any information for the $K^*$ family
in the $K^*\pi$ and $K\rho$ systems. Based on the existing experimental data for the branching fractions of related three-body $B$ decays, PQCD calculations indicate
that the virtual contributions from $K^* \to K^*\pi$ and $K\rho$ in these decays are small. More information on the $K^*$ and its excitations decaying into
$K^*\pi$ and $K\rho$ final states in the $B$ meson decays is expected to be measured in the future high-statistics experiments by Belle II and LHCb.

\begin{acknowledgments}
This work was supported by the National Natural Science Foundation of
China under Grants No. 12205148 and No. 11947011, the Natural Science Foundation of Jiangsu Province under
Grant No. BK20191010, the Qing Lan Project of Jiangsu Province, and the Fund for Shanxi ``1331 Project"
Key Subjects Construction.
\end{acknowledgments}

\appendix
\section{Decay amplitudes}\label{sec-appx-a}
With the subprocesses $K^{*+}\to \{K^{*0}\pi^+, K^0\rho^+$\}, $K^{*0}\to \{K^{*+}\pi^-, K^{+}\rho^-$\}, $K^{*-}\to \{\bar K^{*0}\pi^-, \bar K^{0}\rho^-\}$
and $\bar K^{*0}\to \{K^{*-}\pi^+, K^-\rho^+\}$, the concerned quasi-two-body decay amplitudes are given as follows:
\begin{eqnarray}
{\cal A}\left(B^+\to K^{*0}\pi^+\right)
&=&\frac{G_F}{\sqrt{2}}\big\{V_{ub}^*V_{us}[a_1F^{LL}_{Ah}+C_1M^{LL}_{Ah}]
-V_{tb}^*V_{ts}[(a_{4}-\frac{a_{10}}{2})F^{LL}_{Th}+(C_3-\frac{C_9}{2})M^{LL}_{Th}+(C_5-\frac{C_7}{2})M^{LR}_{Th}\nonumber\\
&&+(a_4+a_{10})F^{LL}_{Ah}+(a_6+a_8)F^{SP}_{Ah}+(C_3+C_9)M^{LL}_{Ah}+(C_5+C_7)M^{LR}_{Ah}]\big\},\label{amp1}
\end{eqnarray}
\begin{eqnarray}
{\cal A}\left(B^+ \to K^{*+}\pi^0\right)
&=& \frac{G_F} {2}\big\{V_{ub}^*V_{us}[a_2F^{LL}_{TK^*}+C_2M^{LL}_{TK^*}+a_1(F^{LL}_{Th}+F^{LL}_{Ah})
+C_1(M^{LL}_{Th}+M^{LL}_{Ah})]-V_{tb}^*V_{ts}[\frac{3}{2}(a_9-a_7)\nonumber\\
&&\times F^{LL}_{TK^*}+\frac{3C_{10}}{2}M^{LL}_{TK^*}+\frac{3C_8}{2}M^{SP}_{TK^*}+(a_4+a_{10})(F^{LL}_{Th}+F^{LL}_{Ah})
+(a_6+a_8)F^{SP}_{Ah}+(C_3+C_9)\nonumber\\
&&\times(M^{LL}_{Th}+M^{LL}_{Ah})+(C_5+C_7)(M^{LR}_{Th}+M^{LR}_{Ah})]\big\},\label{amp2}
\end{eqnarray}
\begin{eqnarray}
{\cal A}\left(B^+ \to K^{*+}\bar{K}^0\right)
&=&\frac{G_F} {\sqrt{2}}\big\{V_{ub}^*V_{ud}[a_1F^{LL}_{AK^*}+C_1M^{LL}_{AK^*}]-V_{tb}^*V_{td}[(a_4-\frac{a_{10}}{2})F^{LL}_{TK^*}
+(a_6-\frac{a_8}{2})F^{SP}_{TK^*}+(C_3-\frac{C_9}{2})\nonumber\\
&&\times M^{LL}_{TK^*}+(C_5-\frac{C_7}{2})M^{LR}_{TK^*}+(a_4+a_{10})F^{LL}_{AK^*}+(a_6+a_8)F^{SP}_{AK^*}+(C_3+C_9)M^{LL}_{AK^*}
+(C_5\nonumber\\
&&+C_7)M^{LR}_{AK^*}]\big\},\label{amp3}
\end{eqnarray}
\begin{eqnarray}
{\cal A}\left(B^+ \to \bar{K}^{*0}K^+\right)
&=&\frac{G_F} {\sqrt{2}}\big\{V_{ub}^*V_{ud}[a_1F^{LL}_{Ah}+C_1M^{LL}_{Ah}]-V_{tb}^*V_{td}[(a_4-\frac{a_{10}}{2})F^{LL}_{Th}
+(C_3-\frac{C_9}{2})M^{LL}_{Th}+(C_5-\frac{C_7}{2})M^{LR}_{Th}\nonumber\\
&&+(a_4+a_{10})F^{LL}_{Ah}+(a_6+a_8)F^{SP}_{Ah}+(C_3+C_9)M^{LL}_{Ah}+(C_5+C_7)M^{LR}_{Ah}]\big\},\label{amp4}
\end{eqnarray}
\begin{eqnarray}
{\cal A}\left(B^0 \to K^{*+}\pi^-\right)
&=& \frac{G_F} {\sqrt{2}}\big\{V_{ub}^*V_{us}[a_1F^{LL}_{Th}+C_1M^{LL}_{Th}]-V_{tb}^*V_{ts}[(a_4+a_{10})F^{LL}_{Th}
+(C_3+C_9)M^{LL}_{Th}+(C_5+C_7)M^{LR}_{Th}\nonumber\\
&&+(a_4-\frac{a_{10}}{2})F^{LL}_{Ah}+(a_6-\frac{a_8}{2})F^{SP}_{Ah}+(C_3-\frac{C_9}{2})M^{LL}_{Ah}
+(C_5-\frac{C_7}{2})M^{LR}_{Ah}]\big\},\label{amp5}
\end{eqnarray}
\begin{eqnarray}
{\cal A}\left(B^0 \to K^{*0}\pi^0\right)
&=& \frac{G_F} {2}\big\{V_{ub}^*V_{us}[a_2F^{LL}_{TK^*}+C_2M^{LL}_{TK^*}]-V_{tb}^*V_{ts}[\frac{3}{2}(a_9-a_7)F^{LL}_{TK^*}
+\frac{3C_{10}}{2}M^{LL}_{TK^*}+\frac{3C_8}{2}M^{SP}_{TK^*}\nonumber\\
&&-(a_4-\frac{a_{10}}{2})(F^{LL}_{Th}+F^{LL}_{Ah})-(a_6-\frac{a_{8}}{2})F^{SP}_{Ah}-(C_3-\frac{C_9}{2})(M^{LL}_{Th}+M^{LL}_{Ah})
-(C_5-\frac{C_7}{2})(M^{LR}_{Th}\nonumber\\
&&+M^{LR}_{Ah})]\big\},\label{amp6}
\end{eqnarray}
\begin{eqnarray}
{\cal A}\left(B^0 \to K^{*+}K^- \right)
&=&\frac{G_F} {\sqrt{2}} \big\{V_{ub}^*V_{ud}[a_2F^{LL}_{AK^*}+C_2 M^{LL}_{AK^*}]-V_{tb}^*V_{td}[(a_3+a_9-a_5-a_7)F^{LL}_{AK^*}
+(C_4+C_{10})M^{LL}_{AK^*}\nonumber\\
&&+(C_6+C_8)M^{SP}_{AK^*}+(a_3-\frac{a_9}{2}-a_5+\frac{a_7}{2})F^{LL}_{Ah}+(C_4-\frac{C_{10}}{2})M^{LL}_{Ah}
+(C_6-\frac{C_8}{2})M^{SP}_{Ah}]\big\},\label{amp7}
\end{eqnarray}
\begin{eqnarray}
{\cal A}\left(B^0 \to K^{*-}K^+ \right)
&=&\frac{G_F} {\sqrt{2}} \big\{V_{ub}^*V_{ud}[a_2F^{LL}_{Ah}+C_2 M^{LL}_{Ah}]-V_{tb}^*V_{td}[(a_3+a_9-a_5-a_7)F^{LL}_{Ah}
+(C_4+C_{10})M^{LL}_{Ah}\nonumber\\
&&+(C_6+C_8)M^{SP}_{Ah}+(a_3-\frac{a_9}{2}-a_5+\frac{a_7}{2})F^{LL}_{AK^*}+(C_4-\frac{C_{10}}{2})M^{LL}_{AK^*}
+(C_6-\frac{C_8}{2})M^{SP}_{AK^*}]\big\},\label{amp8}
\end{eqnarray}
\begin{eqnarray}
{\cal A}\left(B^0 \to K^{*0}\bar{K}^0\right)
&=&-\frac{G_F} {\sqrt{2}}\big\{V_{tb}^*V_{td}[(a_4-\frac{a_{10}}{2})F^{LL}_{TK^*}+(a_6-\frac{a_8}{2})(F^{SP}_{TK^*}+F^{SP}_{AK^*})
+(C_3-\frac{C_9}{2})M^{LL}_{TK^*}+(C_5-\frac{C_7}{2})\nonumber\\
&&\times(M^{LR}_{TK^*}+M^{LR}_{AK^*})+(\frac{4}{3}(C_3+C_4-\frac{C_9+C_{10}}{2})-a_5+\frac{a_7}{2})F^{LL}_{AK^*}
+(C_3+C_4-\frac{C_9+C_{10}}{2})\nonumber\\
&&\times M^{LL}_{AK^*}+(C_6-\frac{C_8}{2})(M^{SP}_{AK^*}+M^{SP}_{Ah})+(a_3-\frac{a_9}{2}-a_5+\frac{a_7}{2})F^{LL}_{Ah}
+(C_4-\frac{C_{10}}{2})M^{LL}_{Ah}]\big\},\label{amp9}
\end{eqnarray}
\begin{eqnarray}
{\cal A}\left(B^0 \to \bar{K}^{*0}K^0\right)
&=&-\frac{G_F} {\sqrt{2}}\big\{V_{tb}^*V_{td}[(a_4-\frac{a_{10}}{2})F^{LL}_{Th}+(a_6-\frac{a_8}{2})F^{SP}_{Ah}
+(C_3-\frac{C_9}{2})M^{LL}_{Th}+(C_5-\frac{C_7}{2})(M^{LR}_{Th}+M^{LR}_{Ah})\nonumber\\
&&+(\frac{4}{3}(C_3+C_4-\frac{C_9+C_{10}}{2})-a_5+\frac{a_7}{2})F^{LL}_{Ah}+(C_3+C_4-\frac{C_9+C_{10}}{2})M^{LL}_{Ah}
+(C_6-\frac{C_8}{2})(M^{SP}_{Ah}\nonumber\\
&&+M^{SP}_{AK^*})+(a_3-\frac{a_9}{2}-a_5+\frac{a_7}{2})F^{LL}_{AK^*}+(C_4-\frac{C_{10}}{2})M^{LL}_{AK^*}]\big\},\label{amp10}
\end{eqnarray}
\begin{eqnarray}
{\cal A}\left(B_s^0 \to K^{*-}\pi^+\right)
&=& \frac{G_F} {\sqrt{2}}\big\{V_{ub}^*V_{ud}[a_1F^{LL}_{TK^*}+C_1M^{LL}_{TK^*}]-V_{tb}^*V_{td}[(a_4+a_{10})F^{LL}_{TK^*}
+(a_6+a_8)F^{SP}_{TK^*}+(C_3+C_9)\nonumber\\
&&\times M^{LL}_{TK^*}+(C_5+C_7)M^{LR}_{TK^*}+(a_4-\frac{a_{10}}{2})F^{LL}_{AK^*}+(a_6-\frac{a_8}{2})F^{SP}_{AK^*}
+(C_3-\frac{C_9}{2})M^{LL}_{AK^*}+(C_5\nonumber\\
&&-\frac{C_7}{2})M^{LR}_{AK^*}]\big\},\label{amp11}
\end{eqnarray}
\begin{eqnarray}
{\cal A}\left(B_s^0 \to \bar{K}^{*0}\pi^0\right)
&=&\frac{G_F} {2}\big\{V_{ub}^*V_{ud}[a_2F^{LL}_{TK^*}+C_2M^{LL}_{TK^*}]-V_{tb}^*V_{td}
[(-a_4-\frac{3a_7}{2}+\frac{5C_9}{3}+C_{10})F^{LL}_{TK^*}-(a_6-\frac{a_8}{2})F^{SP}_{TK^*}\nonumber\\
&&+(-C_3+\frac{3a_{10}}{2})M^{LL}_{TK^*}-(C_5-\frac{C_7}{2})M^{LR}_{TK^*}+\frac{3C_8}{2}M^{SP}_{TK^*}
-(a_4-\frac{a_{10}}{2})F^{LL}_{AK^*}-(a_6-\frac{a_8}{2})F^{SP}_{AK^*}\nonumber\\
&&-(C_3-\frac{C_9}{2})M^{LL}_{AK^*}-(C_5-\frac{C_7}{2})M^{LR}_{AK^*}]\big\},\label{amp12}
\end{eqnarray}
\begin{eqnarray}
{\cal A}\left(B_s^0 \to K^{*+}K^-\right)
&=&\frac{G_F} {\sqrt{2}}\big\{V_{ub}^*V_{us}[a_1F^{LL}_{Th}+C_1M^{LL}_{Th}+a_2F^{LL}_{AK^*}+C_2M^{LL}_{AK^*}]
-V_{tb}^*V_{ts}[(a_4+a_{10})F^{LL}_{Th}+(C_3+C_9)\nonumber\\
&&\times M^{LL}_{Th}+(C_5+C_7)M^{LR}_{Th}+(\frac{4}{3}(C_3+C_4-\frac{C_9+C_{10}}{2})-a_5+\frac{a_7}{2})F^{LL}_{Ah}
+(a_6-\frac{a_8}{2})F^{SP}_{Ah}\nonumber\\
&&+(C_3+C_4-\frac{C_9+C_{10}}{2})M^{LL}_{Ah}+(C_5-\frac{C_7}{2})M^{LR}_{Ah}+(C_6-\frac{C_8}{2})M^{SP}_{Ah}
+(a_3+a_9-a_5-a_7)\nonumber\\
&&\times F^{LL}_{AK^*}+(C_4+C_{10})M^{LL}_{AK^*}(C_6+C_8)M^{SP}_{AK^*}]\big\},\label{amp13}
\end{eqnarray}
\begin{eqnarray}
{\cal A}\left(B_s^0 \to K^{*-}K^+\right)
&=&\frac{G_F} {\sqrt{2}}\big\{V_{ub}^*V_{us}[a_1F^{LL}_{TK^*}+C_1M^{LL}_{TK^*}+a_2F^{LL}_{Ah}+C_2M^{LL}_{Ah}]
-V_{tb}^*V_{ts}[(a_4+a_{10})F^{LL}_{TK^*}+(a_6+a_8)\nonumber\\
&&\times F^{SP}_{TK^*}+(C_3+C_9)M^{LL}_{TK^*}+(C_5+C_7)M^{LR}_{TK^*}+(\frac{4}{3}(C_3+C_4-\frac{C_9+C_{10}}{2})
-a_5+\frac{a_7}{2})F^{LL}_{AK^*}\nonumber\\
&&+(a_6-\frac{a_8}{2})F^{SP}_{AK^*}+(C_3+C_4-\frac{C_9+C_{10}}{2})M^{LL}_{AK^*}+(C_5-\frac{C_7}{2})M^{LR}_{AK^*}
+(C_6-\frac{C_8}{2})M^{SP}_{AK^*}\nonumber\\
&&+(a_3+a_9-a_5-a_7)F^{LL}_{Ah}+(C_4+C_{10})M^{LL}_{Ah}+(C_6+C_8)M^{SP}_{Ah}]\big\},\label{amp14}
\end{eqnarray}
\begin{eqnarray}
{\cal A}\left(B_s^0 \to K^{*0}\bar{K}^0\right)
&=&-\frac{G_F}{\sqrt{2}}\big\{V_{tb}^*V_{ts}[(a_4-\frac{a_{10}}{2})F^{LL}_{Th}+(a_6-\frac{a_8}{2})F^{SP}_{Ah}
+(C_3-\frac{C_9}{2})M^{LL}_{Th}+(C_5-\frac{C_7}{2})(M^{LR}_{Th}+M^{LR}_{Ah})\nonumber\\
&&+(\frac{4}{3}(C_3+C_4-\frac{C_9+C_{10}}{2})-a_5+\frac{a_7}{2})F^{LL}_{Ah}+(C_3+C_4-\frac{C_9+C_{10}}{2})M^{LL}_{Ah}
+(C_6-\frac{C_8}{2})(M^{SP}_{Ah}\nonumber\\
&&+M^{SP}_{AK^*})+(a_3-\frac{a_9}{2}-a_5+\frac{a_7}{2})F^{LL}_{AK^*}+(C_4-\frac{C_{10}}{2})
M^{LL}_{AK^*}]\big\},\label{amp15}
\end{eqnarray}
\begin{eqnarray}
{\cal A}\left(B_s^0 \to \bar{K}^{*0}K^0\right)
&=&-\frac{G_F} {\sqrt{2}}\big\{V_{tb}^*V_{ts}[(a_4-\frac{a_{10}}{2})F^{LL}_{TK^*}+(a_6-\frac{a_8}{2})(F^{SP}_{TK^*}+F^{SP}_{AK^*})
+(C_3-\frac{C_9}{2})M^{LL}_{TK^*}+(C_5-\frac{C_7}{2})\nonumber\\
&&+(M^{LR}_{TK^*}+M^{LR}_{AK^*})+(\frac{4}{3}(C_3+C_4-\frac{C_9+C_{10}}{2})-a_5+\frac{a_7}{2})F^{LL}_{AK^*}
+(C_3+C_4-\frac{C_9+C_{10}}{2})\nonumber\\
&&\times M^{LL}_{AK^*}+(C_6-\frac{C_8}{2})(M^{SP}_{AK^*}+M^{SP}_{Ah})+(a_3-\frac{a_9}{2}-a_5+\frac{a_7}{2})F^{LL}_{Ah}
+(C_4-\frac{C_{10}}{2})M^{LL}_{Ah}]\big\},\label{amp16}
\end{eqnarray}
in which $G_F$ is the Fermi coupling constant and $V_{uq(tq)}$ ($q=b,s$ and $d$) are the CKM matrix elements.
The combinations $a_i$ of the Wilson coefficients are defined as
\begin{eqnarray}
& a_1=C_2+\frac{C_1}{3},\;a_2= C_1+\frac{C_2}{3},\;
\end{eqnarray}
and
\begin{eqnarray}
a_i&=& \left\{\begin{array}{ll}
C_i+\frac{C_{i+1}}{3}, & \ \  {\rm for} \ \ i=3,5,7,9,\\
C_i+\frac{C_{i-1}}{3} & \ \  {\rm for} \ \ i=4,6,8,10.\\
\end{array} \right.
\end{eqnarray}
In Eqs.~(\ref{amp1})$\sim$(\ref{amp16}), the symbols $F$ and $M$ stand for the individual amplitudes of the factorizable and
nonfactorizable Feynman diagrams.
The superscripts $LL$, $LR$, and $SP$ denote the contributions from $(V-A)(V-A)$, $(V-A)(V+A)$, and $(S-P)(S+P)$ operators, respectively.
The subscripts $TK^*$ and $Th$ identify the amplitudes from the emission diagrams in Figs.~\ref{fig-feyndiag}(a) and ~\ref{fig-feyndiag}(c),
 corresponding to the $B\to K^*$ and $B\to h$ transitions, while the subscripts $AK^*$ and $Ah$ refer to the amplitudes for the
annihilation diagrams in Figs.~\ref{fig-feyndiag}(b) and~\ref{fig-feyndiag}(d).
The specific expressions for these general amplitudes are the same as in the appendix of Ref.~\cite{epjc80-815} with the replacement
$\phi \to K^*$, and are omitted here.


\end{document}